\documentclass[11pt,twoside]{article}
\usepackage{amsfonts}
\usepackage{amssymb,latexsym,amsmath}
\usepackage{graphicx}
\usepackage[mathscr]{eucal}

\newcommand\aut{\operatorname{Aut}}

\def\basisp #1 #2{{\frac{\partial}{\partial p^{#1}_{#2}}}}
\def\basispi #1 #2{{\frac{\partial}{\partial \pi^{#1}_{#2}}}}

\def\basisx #1{{\frac{\partial}{\partial x^{#1}}}}
\def\basisy #1{{\frac{\partial}{\partial y^{#1}}}}
\newcommand\blob{\mbox{$\;\Box$}}

\newcommand\define{\bigskip\noindent{\bf Definition}$\;\;$} 
\newcommand\diff{\operatorname{Diff}}

\newcommand\ga{{G_{\cal A}}}

\newcommand\hook{\,\hbox to 10pt{\vbox{\vskip 6pt\hrule width 6.5pt height 1pt}
         \kern -4.0pt\vrule height 8pt width 1pt\hfil}\,}
\def\iwtheta #1{{\wedge^{#1}i^*\theta}}          
\def\lb2{[\![}
\newcommand\lie{\mbox{\textbf{L}}}
\newcommand\lieG{\mbox{$\mathscr{G}$}}
\newcommand\lieg{\mbox{$\mathfrak g$}}
\newcommand\lieh{\mbox{$\mathfrak h$}}

\newcommand\lvy{{L_VY}}

\newcommand\nka{ \left( \begin{array}{cc} N&0\\ A&K \end{array}\right)}

\newcommand\ok{O_\iota(k)}
\newcommand\okalg{\mathfrak{o}_\iota(k)}
\newcommand\on{O_\eta(n)}
\newcommand\onalg{\mathfrak{o}_\eta(n) }
\newcommand\onk{\on\times \ok}
\newcommand\onkalg{\onalg \times \okalg}
\def\partialq #1 #2{{\frac{\partial #1}{\partial q^{#2}}}}
\def\partialx #1 #2{{\frac{\partial #1}{\partial x^{#2}}}}
\def\partialy #1 #2{{\frac{\partial #1}{\partial y^{#2}}}}
\def\pb#1#2{\{\hat #1,\hat #2\}}
\newcommand\phibl{{\phi_{(B,\lambda)}}}

\newcommand\projectable{{\cal X}_{\mbox{\tiny Proj}}Y}
\newcommand\proof{\noindent{\bf Proof}$\;\;$}      
\newcommand\ptlvy{{(y,\{e_i,\epsilon_A\})}}
\def\r{\mbox{$\mathbb R$}}
\def\rb2{]\!]}
\newcommand\rem{\bigskip\noindent{\bf Remark}$\;\;$}  
\newcommand\rkn{{\r^{k\times n}}}
\newcommand\rnk{{\r^{n\times k}}}

\newcommand\rnplusk{\r^{n+k}}
\def\stvlvy{ST^2_V(\lvy)}
\newcommand\tr{\operatorname{tr}}
\newcommand\tvlvy{T^1_V(\lvy)}

\def\xhat #1{{X_{\hat {#1}}}}

\newtheorem{thm}{Theorem} [section]
\newtheorem{lemma}[thm]{Lemma}           

\newtheorem{prop}[thm]{Proposition}

\begin{document}

\title{
A Frame Bundle Generalization of 
Multisymplectic Momentum Mappings
\thanks{{\em 2000 MSC:} 53 D 20, 70 G 45, 57 R 15, 53 C 15, 37 J 15} 
\thanks{{\em Keywords:} multisymplectic geometry,
frame bundle, covariant field theories, Poisson bracket, 
momentum mappings.
}
\author{
J. K. Lawson\\
Department  of Mathematics,
Trinity University,
715 Stadium Drive, \\
San Antonio, TX 78212-7200
(e-mail: jlawson@trinity.edu)
\\ 
\\ 
\date{}
}}
\maketitle

\begin{abstract} 
This paper presents
generalized momentum mappings for covariant Hamiltonian 
field theories.
The new momentum mappings arise from 
a generalization of symplectic geometry to $\lvy$, the bundle of
vertically adapted linear frames over the bundle
of field configurations $Y$.
Specifically, the generalized field momentum observables are
vector-valued momentum mappings on the vertically adapted frame 
bundle generated from automorphisms of $Y$.
The generalized symplectic geometry on $\lvy$ is 
a covering theory for multisymplectic geometry on the
 multiphase space $Z$, 
and it follows that the field momentum observables
on $Z$ are generalized by those on $\lvy$.
Furthermore, momentum observables on $\lvy$ produce 
conserved quantities along flows in $\lvy$.
For translational and orthogonal symmetries of fields and reparametrization symmetry in mechanics,
momentum is conserved,
and for 
angular momentum 
in time-evolution mechanics 
we produce a 
version of the parallel axis theorem of rotational dynamics,
and in special relativity, we produce the transformation of angular momentum under boosts. 
\end{abstract}

\thispagestyle{empty}

\tableofcontents

\section{Introduction}

\setcounter{equation}{0}

Norris' generalization~\cite{No1,No2,No3} of the symplectic geometry 
of the cotangent bundle 
has been a successful theoretical tool for particle mechanics.
Norris has shown that the standard symplectic
geometry of the cotangent bundle to an $n$-dimensional manifold 
may be
obtained in its entirety from the {\em $n$-symplectic} geometry
of the linear frame bundle.  The key component of $n$-symplectic
geometry is recognizing that the canonical soldering one-form
of a frame bundle~\cite{KN} may be employed as a vector-valued $n$-symplectic
potential.

Subsequently, it has been shown~\cite{FLN2,La} that 
the multisymplectic geometry of the 
(affine) multiphase space $Z$ introduced by
Kijowski~\cite{Ki1,Ki2} and refined by Gotay, et al.~\cite{GIMMsy},
may be generalized by adapting Norris's
theory to $\lvy$, the bundle of vertically adapted linear frames of a
configuration bundle $Y$ of a classical field.
We may use the
 new theory to reproduce the Poisson  bracket of momentum observables  on $Z$.
However,  the momentum observables on $Z$ are not closed under this  bracket, whereas
the analogous momentum
observables on $\lvy$ are closed under the analogous Poisson bracket~\cite{FLN2,La}.

The purpose of this paper is to extend this work in generalized symplectic
geometry to include momentum mappings for field theories. In the 
Hamiltonian framework, momentum mappings are the foundation for 
obtaining
conserved quantities from group symmetries of phase space (the symplectic or 
multisymplectic manifold). 
Because the generalized symplectic structure is
invariant under  automorphisms of $Y$ lifted to $\lvy$, 
 momentum mappings
on $\lvy$ induce the momentum mappings on $Z$ found in~\cite{GIMMsy}.
This is analogous to the result of Norris~\cite{No2}, which induces the momentum
mappings of standard symplectic geometry of the cotangent bundle from
the theory of $n$-symplectic geometry.
The momentum observables on $\lvy$ established in~\cite{FLN2} and~\cite{La}
are shown to be a special case of the new momentum mappings 
obtained from the Lie algebra of projectable vector fields on $Y$.

When we consider momentum mappings constructed from the infinitesimal
generators of the translation group, an orthogonal group, or an 
affine group, we obtain conserved quantities
along flows that give us ``frame bundle'' versions of
conservation of linear field momentum, conservation of angular field momentum,
and a ``parallel axis theorem'' for field theories.
We may also use time-reparametrization symmetry in mechanics to extend an
application of Norris~\cite{No1,No3} regarding parallel transport of frames
along geodesics of a Riemannian metric.

The format of this paper is as follows.  First we summarize multisymplectic geometry in Section 2, 
and then we introduce momentum mappings on $\lvy$ in Section 3.  
In Section 4 we generate momentum mappings on $Z$ from those on $\lvy$,
and in Section 5 we derive conserved quantities.  Examples are found in Section 6.

\section{Momentum mappings in multisymplectic geometry}
\label{sec:multisymplectic}
\setcounter{equation}{0}

Let $X$ be an oriented $n$-dimensional manifold   
and let ${\pi_{XY}}: Y \rightarrow X$ be a fiber bundle
 with standard fiber a $k$-dimensional manifold.  
(Note:  In general, we shall denote a 
projection from $A$ onto $B$ as $\pi_{BA}$.)
A classical field is a section of 
the {\em field configuration space}
$Y$ over the {\em parameter space} $X$.
From local coordinates $\{x^i\}$, $i = 1,\dots,n$ on $X$ we may construct local adapted coordinates $\{ x^i, y^A\}$, $i = 1,\dots,n$, $A = 1,\dots,k$, 
on $Y$.
The \textit{multivelocity} 
bundle is the (first-order) {\em jet bundle} $JY$,
the affine bundle
over $Y$ whose fiber over $y\in Y$ consists of linear maps 
$\gamma_y: T_{{\pi_{XY}}(y)}X\rightarrow T_yY$ satisfying
${\pi_{XY}}_*\circ \gamma_y = \mbox{Id}\,_{T_{{\pi_{XY}}(y)}X}$.
A section of $JY$
over $Y$ can be identified
 with an Ehresmann connection
 on $\pi_{XY}:Y\rightarrow X$.

The {\em  bundle of affine cojets}~\cite{Go,GIMMsy} $J^\star Y$ is the vector
bundle over $Y$ whose fiber at $y$ is the set of affine maps from $J_yY$ to 
$\wedge^{n}_{{\pi_{XY}}(y)}X$.
It follows that  
$\dim J^\star Y  = \dim JY + 1$.
An equivalent
description~\cite{GIMMsy} of $J^\star Y$ is useful.
Define the {\em vertical subbundle} of $TY$ to be 
$$V(TY) := \{ w_y\;|\; y\in Y , \, w_y \in T_yY \, \mbox{and}
\; \pi_{XY\,*}(w_y)=0 \}\, .$$  
The {\em multiphase space} $Z$~\cite{GIMMsy,Ki2} 
is the fiber bundle whose 
fiber $Z_y$ 
over $y \in Y$ is 
$$Z_y := \{ z\in \wedge^n_yY  \, | \, v\hook w\hook z = 0 \; \forall v,\, w \in
 V(T_yY)\}\, ,$$
where $\hook$ denotes the inner product of a vector with a differential form.
The bundle $Z$, originally defined by Kijowski~\cite{Ki2}, 
admits a canonical
$n$-form, 
\[
\Theta(z) = \pi^*_{YZ}(z) \, , 
\]
which is the pullback via inclusion 
$Z \hookrightarrow \wedge^nY$
of the canonical $n$-form on $\wedge^nY$,
and $Z$ is 
``canonically'' isomorphic to $J^\star Y$~\cite{GIMMsy}.
We can define coordinates $\{ x^i, y^A, p^j_B, p\}$ on $Z$ 
where $\{x^i,y^A\}$ are the lifts of the
 adapted coordinates of $Y$, 
\begin{eqnarray}
  p(z) & = &  \basisx n \hook\cdots\hook\basisx 1\hook z \, , 
\quad\mbox{and} \label{eqn:zcoords}\\
p^j_B(z) & = & (-1)^{j-1} \, \basisx n \hook \basisx {n-1} \hook\dots\hook 
\widehat{\basisx j} \hook \dots\hook\basisx 1\hook\basisy B \hook z \, ,
\nonumber 
\end{eqnarray}
where $\widehat{\basisx j}$ denotes the omission of ${\basisx j}$.
If we define in local lifted coordinates
\[
d^nx :=  dx^1 \wedge \cdots \wedge dx^n \, , \;\;
d^{n-1} x_i  : =  \basisx i \hook d^nx \, , \;\;  	
\mbox{and} \;\;
d^{n-2} x_{ij}   : =  \basisx j \hook \basisx i \hook d^nx  \, ,
\]
then $\Theta$ may be expressed locally as
\[
\Theta = p^i_A dy^A \wedge d^{n-1}   x_{i} +  p d^nx
\]
The $(n+1)$--form $d\Theta$ is nondegenerate and thus may be
considered a \textit{multisymplectic structure form} on $Z$.
The pair $(Z,d\Theta)$ is 
called a 
{\em multisymplectic manifold}~\cite{GIMMsy}.

\bigskip\noindent{\bf Definition}~\cite{Go,GIMMsy}
 Let a Lie group $\lieG$  act on the left on $Z$ and
 the multisymplectic structure form $d\Theta$
be invariant under this action.  
Let $\lieg$ be the Lie algebra of $\lieG$.
 A mapping 
$$J: Z\rightarrow \lieg^*\otimes \wedge^{n-1}Z$$ is a 
{\em momentum mapping} if $J$
covers the identity on $Z$ and if 
$\hat J$  is defined by
$$\hat J(\xi)(z) = \left< J(z),\xi\right>$$
then 
$$
d\hat J (\xi) = -\xi_Z\hook d\Theta \, ,
$$
where $\xi_Z$ is the infinitesimal generator of the $\lieG$-action 
on $Z$ induced
by $\xi \in\, \lieg$.

\define A {\em Hamiltonian observable} on $Z$ is an 
$(n-1)$-form $f$ on $Z$ that satisfies 
\begin{equation}\label{eqn:Hamobservable}
df = - X \hook d\Theta
\end{equation}
for some vector field $X$ on $Z$.
The vector field $X$ is called a {\em Hamiltonian vector field} on $Z$.

\bigskip\noindent
For a momentum mapping $J$ and $\xi \in \lieg$,
$\hat J (\xi)$ is a Hamiltonian observable and
$\xi_Z$ is the corresponding Hamiltonian vector field.

The group $\aut Y$ of fiber bundle automorphisms of $Y$ over $X$
is the multisymplectic analogue of the
group $\diff X$ of diffeomorphisms of configuration space in 
the symplectic geometry of the cotangent bundle $T^*X$.
Let ${\cal X}Y$ be the Lie algebra of  vector fields on $Y$.
Denote the space of
vector 
fields of $Y$ projectable to $X$ by $\projectable$.
Note that $\projectable$ is a Lie subalgebra
of ${\cal X}Y$, since $[{\pi_{XY}}_*v,{\pi_{XY}}_*w] = {\pi_{XY}}_*[v,w]$~\cite[p.\ 85]{Ch}.

\begin{lemma}~The group $\aut\, Y$ formally has for its Lie algebra 
the vector space of complete projectable vector fields on $Y$.
\end{lemma}

\proof (Motivated by~\cite{AM}.)
We identify the Lie algebra of $\aut Y$ with the set of all vector fields
of $Y$ which are tangents to curves through the identity of $\aut Y$.
Let $\{ {\eta_Y}_\lambda \, |\, \lambda \in \r\}$ be a curve in $\aut Y$ such that 
${\eta_Y}_0 = \mbox{Id\,}_Y$. 
For
each ${\eta_Y}_\lambda$ there exists ${\eta_X}_\lambda$,
 a diffeomorphism of $X$ such that ${\pi_{XY}}\circ {\eta_Y}_\lambda = {\eta_X}_\lambda \circ {\pi_{XY}}$.  Since ${\eta_Y}_\lambda \in\, \aut Y \subset \diff Y$, 
 $v := \left. {\frac{d}{d\lambda}}{\eta_Y}_\lambda \right| {}_{\lambda = 0}$ is a tangent vector at $\mbox{Id\,}_Y$,
$\tau_Y\circ v = \mbox{Id\,}_X$
 and 
\[
{\pi_{XY}}_*\circ v(y) =  
\left. \left. \frac{d}{d\lambda}({\pi_{XY}}\circ \eta_\lambda(y))\,\right|_{\lambda = 0} 
=    \frac{d}{d\lambda}{\eta_X}_\lambda\,\right|_{\lambda = 0}({\pi_{XY}}(y)) 
=:   {\underline v} \circ{\pi_{XY}}(y)\; .
\]
Observe that if $\pi_{X\,TX}:  TX \rightarrow X$ is  projection then
 $\pi_{X\,TX}\circ \underline v = \mbox{Id}_X$.
Therefore, $T_{\mbox{\tiny Id}_Y}\aut Y$
is isomorphic to a vector subspace of $\projectable$.
In the case that $Y$ is compact we shall
 show that $T_{\mbox{\tiny Id}_Y}\aut Y$ is all of the space $\projectable$.
 Let $\phi$ be the flow box~\cite{AM} for $v$ at $y$ and 
 let $\underline \Phi$
be the  flow box for $\underline v$ at ${\pi_{XY}}(y)$.
  Then 
for each $y \in Y$, we have 
 that
${\pi_{XY}}\circ\Phi_0(y) = \underline\Phi_0\circ{\pi_{XY}} (y)$, 
$$
 \frac{d}{dt}({\pi_{XY}}\circ\Phi_t)(y) = {\pi_{XY}}_* v(\Phi_t(y)) 
= \underline v \left( {\pi_{XY}} (\Phi_t(y)) \right) = \underline v \left(({\pi_{XY}}\circ
\Phi_t)(y)\right)
$$
and 
$$
 \frac{d}{dt}(\underline\Phi_t \circ {\pi_{XY}})(y) = 
\underline v\left(\underline \Phi_t( {\pi_{XY}}(y))\right) = 
\underline v\left((\underline \Phi_t\circ {\pi_{XY}})(y)\right)\, .
$$
Thus 
${\pi_{XY}}\circ\Phi_t(y)$ and $\underline\Phi_t\circ{\pi_{XY}} (y)$
are curves in $X$ that solve the same initial value problem on  open 
neighborhoods  of $t$ and  of ${\pi_{XY}}(y)$.  By an
existence and uniqueness theorem~\cite{AM}, they are equal.
Therefore 
${\pi_{XY}}\circ\Phi_t = \underline\Phi_t\circ{\pi_{XY}}$ so the 
projectable vector field $v$
defines a one-parameter curve $\{\Phi_t \, | \, t \in \r\}$ in $\aut Y$ such that
$v = \frac{d}{dt} \Phi_t |_{t=0}$.~\blob

\bigskip
\noindent We shall  formally identify ${\cal X}Y$ with the tangent space at the identity of
$\diff Y$, the group of diffeomorphisms of $Y$.  
Subsequently, we will formally identify $\projectable$ 
with the tangent space at the identity of $\aut Y$.
However this ignores the topology of $Y$.
Milnor~\cite{Mi} notes that if $Y$ is compact then $\diff Y$
is a Lie group that admits a $C^\infty$ manifold structure. 
For noncompact $Y$ we may choose only to consider diffeomorphisms which are
the identity outside a compact subset of $Y$.  
This leads to vector fields that vanish outside this subset.  

There is a slight ambiguity because $\projectable$ serves both as the Lie algebra and as the collection of infinitesimal generators. To resolve this,
let $[\xi,\zeta]$ denote the Lie bracket defined on the formal Lie algebra of left invariant vector fields on the manifold $\aut Y$, 
and let $[\xi_Y,\zeta_Y]$ denote the usual Lie bracket defined on ${\cal X}Y$ (in which the vector fields are right invariant with 
respect to the $\aut Y$ action).  
See~\cite[Exercise 4.1G]{AM} for a clarification.

The {\em canonical lift}~\cite{GIMMsy} of $\eta_Y \in \aut Y$ is a map 
\[
\eta_Z : Z \rightarrow Z
\; : \;
z \mapsto (\eta_Y^{-1})^*(z) \, .
\]
The map $\eta_Z$ is a 
$\pi_{XZ}$--bundle map 
under which $\Theta$ is invariant. 
For $\xi \in \projectable$, it follows that $\lie_{\xi_Z}\Theta = 0$
and the induced momentum mapping defines the \textit{momentum observable}
\begin{equation}\label{eqn:specialmommap}
\hat J(\xi)(z) := \xi_Z \hook \Theta (z) = \pi_{YZ}^* (\xi_Y \hook z)\, ,
\end{equation}
where $\xi_Y$ is the infinitesimal generator of the ensuing action of
$\aut Y$ on $Y$.
It follows that $\hat J$ is $Ad^*$-equivariant.
That is,  $J(Ad^{-1}_\eta \xi) = \eta_Z^*(\hat J(\xi))$.
Let $T^1(Z)$ denote the vector space of momentum observables,
and
observe that $\xi \mapsto \hat J(\xi)$ is a bijection from $\projectable$ to $T^1(Z)$.

If in local adapted coordinates on we write 
$\xi \in \projectable$ as
$\xi = \xi^i(x^j)\basisx i + \xi^A(x^j,y^B)\basisy A$, 
then 
	\begin{equation}\label{eqn:fv}
\hat J(\xi) (z)
 =  (p^i_A\xi^A + p\xi^i)d^{n-1}x_i - p^i_A \xi^j dy^A\wedge d^{n-2}x_{ij} 
	\end{equation}
and 
	\begin{equation} \label{eqn:xfvcoords}
\xi_Z  = \xi^k \basisx k + \xi^A \basisy A 
+  \left( p^j_A \partialx {\xi^i} j - p^i_A\partialx {\xi^j} j
- p^i_B \partialy {\xi^B} A \right)   \basisp i A
-  \left( p\partialx {\xi^i} i  + p^i_A\partialx {\xi^A} i  \right)   
\basisp {} {}\, .
	\end{equation}
From (\ref{eqn:xfvcoords}) if 
$\xi_Y,\zeta_Y \in \projectable$ then 
$\pi_{YZ*}\left[\xi_Z,\zeta_Z \right] =  [\xi_Y,\zeta_Y]$,
and thus, by~(\ref{eqn:specialmommap}),
\begin{equation}	
\left[\xi_Z,\zeta_Z \right] \hook\Theta_z 
 =  \pi^*_{YZ}\left( [\xi_Y,\zeta_Y]\hook z\right) \, . 
\end{equation}

If we define a Poisson bracket by
\begin{equation}\label{eqn:pb}
\{\hat J(\xi), \hat J(\zeta)\} := - d\Theta(\xi_Z,\zeta_Z)
\end{equation}
then using $\lie_{\xi_Z}\Theta = 0$ and  the
Lie derivative identities 
\cite[p.\ 121]{AM},
	\begin{equation}\label{eqn:liederiv}
[X,Y]\hook \alpha = \lie_X(Y\hook \alpha) - Y\hook(\lie_X\alpha)\, 
\quad\mbox{and}\quad
\lie_X\alpha = X \hook d\alpha + d(X\hook \alpha) \, ,
	\end{equation}
we obtain
\begin{equation}\label{eqn:Jbrackets}
\{\hat J(\xi),\hat J(\zeta)\} =   \hat J([\xi,\zeta]) -  d(\xi_Z\hook \zeta_Z \hook \Theta) \, .
\end{equation}
For an alternative proof using the $\operatorname{Ad}^*$ equivariance of $J$, see~\cite[Prop.\ 4.5]{GIMMsy}.
 
The  Poisson bracket on $T^1(Z)$
given by~(\ref{eqn:pb}) 
is not a true Poisson bracket 
because there lacks an 
associative multiplication of $(n-1)$--forms on which the bracket acts
as a derivation.
Worse, $d(\xi_Z\hook \zeta_Z \hook \Theta)$ is not 
in $T^1(Z)$
because from equation~(\ref{eqn:Hamobservable})
the Hamiltonian vector field of an exact form is the zero vector field on $Z$, but
the momentum observable corresponding to the zero vector field is
the zero $(n-1)$-form on $Z$.
Thus, $T^1(Z)$
is {\bf not} closed under the Poisson bracket.
Equation (\ref{eqn:Jbrackets}) explains the remark in~\cite{GIMMsy}
 that the
Poisson bracket of two momentum observables ``is up to the addition of
exact terms, another momentum observable.''

\section{Momentum mappings on the vertically adapted linear
frame bundle}
\setcounter{equation}{0}

For an $n$-dimensional manifold $M$, 
the linear frame bundle $LM$ has an $\r^n$-valued canonical soldering one-form~\cite{KN},
which we regard as an 
\textit{$n$-symplectic} potential.  
This geometric object has been shown~\cite{No2}
to generate the
canonical one-form on the cotangent bundle $T^*M$. 
Norris's program
of $n$-symplectic geometry~\cite{No1,No2,No3} originates
from the
exterior derivative of the soldering form, which is a closed nondegenerate
two-form serving as an
$\r^n$-valued \textit{$n$-symplectic structure} form.
The $n$-symplectic geometry 
generalizes the symplectic geometry of classical particle mechanics on $T^*M$.

Now, for the $(n+k)$--dimensional field configuration bundle $Y$, 
we shall treat the canonical soldering form $\theta$ on
$LY$ as an $(n+k)$-symplectic potential  $\theta$.

\define (motivated by the definition of an {\em adapted frame}~\cite{KN})
The {\em vertically adapted frame bundle} $\lvy$ 
is defined to be
\[
\lvy  := \{\ptlvy \in LY \;|\; 
\{\epsilon_A\} \hbox{ is a frame of } V(T_yY) \}\, .
\]

\medskip

\noindent The bundle $\lvy$ is a reduced subbundle of $LY$ obtained by breaking
the $GL(n+k)$ symmetry of $LY$~\cite{La}.

The structure group of $\lvy$,
expressed with respect to
the standard basis of $\rnplusk$ is the {\em adapted linear group} 
	\[ 
\ga  := \left\{\nka \left. 
\;|\; 
N \in GL(n),\, K \in GL(k),\, A \in 
{\rkn} \right. \right\} \, ,
	\]
with multiplication defined by block matrix multiplication.
For convenience we write $\nka\in \ga$ as $(N,K,A)$.
The free right action of $\ga $ on $\lvy$ is given by 
\begin{equation}
\ptlvy \cdot (N,K,A) := (y,\{e_jN^j_i+\epsilon_BA^B_i,\epsilon_BK^B_A\}) \,.
\label{eqn:galvy}
\end{equation}
Let $(y,\{e^i, \epsilon^A\})$ be the coframe dual to the vertically
adapted frame $\ptlvy$.
We may define the right action of $\ga $ on 
$(y,\{e^i,\epsilon^A\})$ by   
\begin{equation}
(y,\{e^i,\epsilon^A\})\cdot(N,K,A)
 =  (y,\{(N^{-1})^i_je^j,-(K^{-1}AN^{-1})^A_je^j  + (K^{-1})^A_B\epsilon^B\})\, . 
\label{eqn:coframeaction}
\end{equation}
The  coframe in~(\ref{eqn:coframeaction}) is dual to the frame
$\ptlvy\cdot(N,K,A)$.

The pullback of the   $(n+k)$-symplectic structure form $d\theta$ 
via  inclusion $i:\lvy\hookrightarrow LY $
is closed and  nondegenerate on $\lvy$,
 just as $d\theta$ is on
$LY$. 
Local coordinates on $\lvy$ are  $\{x^i,\,y^A,\,\pi^i_j,\,\pi^A_B,\pi^A_i\}$,
where 
$\{x^i, y^A\}$ are local adapted coordinates on $Y$
and
$ \pi^i_j := e^i(\basisx j)$, $ \pi^A_B := \epsilon^A(\basisy B)$ and 
$ \pi^A_j := \epsilon^A(\basisx j)$.
Let $\{R_\mu\}_{\mu=1,2,\dots,n+k}$ be the standard basis of $\rnplusk$,
$\{r_i\}_{i=1,2,\dots,n}$ be the standard basis of $\r^n$, and
$\{s_A\}_{A=1,2,\dots,k}$ be the standard basis of $\r^k$.
Then,
we define
$\hat r_i := (r_i,0) \in \rnplusk$
and
$\hat s_A := (0,s_A) \in \rnplusk$.
Thus, in local coordinates,
\[
i^*\theta = \pi^i_j dx^j \hat r_i + 
(\pi^A_i dx^i +  \pi^A_B dy^B )\hat s_A 
\]
and
\begin{equation}\label{eqn:idtheta}
i^*d\theta = d\pi^i_j\wedge dx^j \otimes \hat r_i + 
(d\pi^A_i\wedge dx^i +  d\pi^A_B \wedge dy^B )\otimes\hat s_A \, .
\end{equation}

\define A \textit{Hamiltonian observable} on $\lvy$ is an 
$\r^n$-valued function $\hat f$ on $\lvy$ that satisfies
\begin{equation}
d\hat f = -X_{\hat f}\hook i^*d\theta\; ,
\label{eqn:nkstruc}
\end{equation}
for some
$X\in{\cal X}(\lvy)$.
The vector field $X_{\hat f}$ is then called a \textit{Hamiltonian vector field} on $\lvy$.

\bigskip\noindent As in the $n$-symplectic theory on 
$LM$~\cite{No1,No2,No3}, 
equation~(\ref{eqn:nkstruc}) admits neither all vector fields
nor all $\rnplusk$-valued functions.  

\bigskip\noindent
{\bf Definitions} \begin{description}
\item[{$\bullet$}]  The vector space of {\em Hamiltonian observables} on $\lvy$
is denoted by $HF^1(\lvy)$.
\item[{$\bullet$}] The vector space of  {\em Hamiltonian vector fields} on $\lvy$
 is denoted by $HV^1(\lvy)$.
\item[{$\bullet$}] $T^1(\lvy)$ is the vector space of tensorial $\r^{n+k}$-valued functions on $\lvy$.
\end{description}

\noindent  
An element of $T^1(\lvy)$ can be expressed in local coordinates as 
\[
\hat f = f^i(x^k,y^C)\pi^j_i\otimes\hat r_j 
+ f^A(x^k,y^C)\pi^B_A\otimes\hat s_B
+ f^i(x^k,y^C)\pi^B_i\otimes\hat s_B \, .
\]
For $\hat f \in T^1(\lvy)$ we solve 
equation~(\ref{eqn:nkstruc}) 
locally  for $X_{\hat f}$.   This yields
\[
X_{\hat f}  =  f^i\basisx i + f^A\basisy A   
- \partialx {f^k} j \pi^i_k\basispi i j  -  \partialy {f^C} B \pi^A_C
\basispi A B
-\left( \partialx {f^j} i \pi^A_j + \partialx {f^B} i \pi^A_B  \right)
\basispi A i \, ,    
\]
subject to the  constraints on $\hat f$,
\begin{equation}\label{eqn:constraint}
\partialy f^i A  = 0 \quad
\forall i = 1,\dots,n \;\;\mbox{and}\;\; A= 1,\dots,k.
\end{equation}
Thus, $T^1(\lvy)\not\subset HF^1(\lvy)$. 
Define
$T^1_V(\lvy) := HF^1(\lvy)\cap  T^1(\lvy)$.
Since $\lvy$ is a subbundle of $LY$,
a point $w\in\,\lvy$ is a linear isomorphism~\cite{KN}
$w:\r^{n+k}\rightarrow T_{\pi_{Y\,\lvy}(w)}Y\, .$
Moreover, for $g \in \ga , (w\cdot g)V = w (g V)$ for each $V\in \rnplusk$.
Hence a tensorial function $\hat f$ 
corresponds bijectively to
$\mathbf{f}\in\,{\cal X}Y$ by the relation $\hat f(w) = w^{-1}(\mathbf{f}
 (\pi_{Y\,\lvy} (w)))$.
Now, if $\hat f\in\,T^1_V(\lvy)$ then,  by~(\ref{eqn:constraint}), $f^i = f^i(x^j)$ and
$\hat f$ is induced from a projectable vector field. 
Conversely, $\mathbf{f} \in \projectable$ gives a tensorial function $\hat f$
satisfying~(\ref{eqn:constraint}).
Combine this with the results of Section 2 to obtain the
following proposition~\cite{La}.  
\begin{prop}\label{prop:bijective}
$\;T^1_V(\lvy)$, $T^1(Z)$, and $\projectable$ are in pairwise bijective correspondence.
Furthermore, if  $n \geq 2$ and $k \geq 2$ then
$HF^1(\lvy) \simeq T^1_V(\lvy)\oplus C^\infty(X,\r^n)\oplus C^\infty(Y,\r^{k})$.
\end{prop}

\define
Let $\lieG$ be a Lie group with an  action 
 on $\lvy$ under which $i^*d\theta$ is invariant.  
Let $\lieg$ be the Lie algebra of $\lieG$.
A mapping $J:\lvy\rightarrow \lieg^*\otimes\r^{n+k}$ is a {\em momentum mapping} if 
\[
d\hat J (\xi) = -\xi_\lvy\hook di^*\theta\quad \forall\, \xi\in\, \lieg
\]
where $\xi_\lvy$ is the infinitesimal generator of the $\lieG$--action  
on $\lvy$ generated by $\xi \in \lieg$ and
$\hat J (\xi) : \lvy \rightarrow \rnplusk$ is defined by
\[
\hat J(\xi)(w) = \; <J(w),\xi>\quad\forall\, w\in\,\lvy\, .
\]

\medskip\noindent  
Note that a momentum mapping on $\lvy$ is the restriction 
of a momentum mapping on $LY$ (as defined in~\cite{No2}).
When necessary, we will write $J=J_{\lvy}$ to clarify that $J$ 
is a momentum mapping on $\lvy$.
Furthermore, for any $\xi \in \lieg$,
$\hat J(\xi) \in HF^1(\lvy)$ and its corresponding Hamiltonian vector field $X \in HV^1(\lvy)$.

Now we consider momentum mappings when $\lieG = \aut Y$.
Define $\aut(\lvy)$ to be the 
 subgroup of $\diff(\lvy)$
whose elements are fiber bundle automorphisms 
both over $Y$ and over $X$.
Let $\eta_Y\in \aut Y$ cover $\eta_X \in \diff X$.
Define the mapping 
\[
\eta_{\lvy}: \lvy\rightarrow\lvy\; :\; \ptlvy\mapsto
(\eta_Y(y), \{ \eta_{Y*}e_i, \eta_{Y*}\epsilon_A\} )\, .
\]
Clearly, $\eta_{\lvy}$ is bijective and smooth and has a smooth
inverse.
By definition, $\pi_{Y\,\lvy} \circ \eta_{\lvy} = \eta_Y \circ \pi_{Y\,\lvy}$.
Since  ${\pi_{XY}}_*\eta_{Y*}\epsilon_A = \eta_{X*}{\pi_{XY}}_*\epsilon_A = 0$,
it follows that $\eta_{Y*}\epsilon_A\in\,V(T_{\eta_Y(y)}Y)$ and
 thus $\eta_{\lvy} \in \,\aut(\lvy)$.  
Therefore,
 $\eta_\lvy$ is the
\textit{canonical lift} of  
$\eta_Y$.

\rem We cannot naturally define a lift of  $\diff Y$ to $\aut(\lvy)$.
Indeed, let $X= \r$ and $Y = X\times \r$.
Define $f\in\diff Y$ 
by $f(x,y) = (y,x)$.  
Then   
$f_*(\frac{\partial}{\partial y}) = \frac{\partial}{\partial x}$,
which implies that $f_*(V(TY)) \not\subseteq V(TY)$.
  
\begin{prop}\label{thm:liftlvy}~The $(n+k)$--symplectic potential $i^*\theta$ on $\lvy$ is invariant under  the canonical lift of each automorphism of $Y$.
Conversely, every element of $\aut(\lvy)$ that leaves ${i^*\theta}$ invariant
is a lift of an automorphism
of $Y$. 
\end{prop}

\proof 
We modify a result of Kobayashi and Nomizu~\cite{KN}.
Let $\eta_Y \in \aut Y$ and let $\eta_\lvy$ be its lift.
We may view any $w\in \lvy$ as a 
linear isomorphism $w:\rnplusk \rightarrow T_{\pi_{Y\,\lvy(y)}}Y$.
Then, as linear isomorphisms,
$\eta_{\lvy}(w) = \eta_{Y*w}\circ w $,
and thus $\eta_{\lvy} (w)^{-1}\circ \eta_{Y*w} = w^{-1}$.  
So, for $X_w \in T_w(\lvy)$,
\begin{eqnarray*}
\eta_{\lvy}^* i^*\theta(X_w) & = & i^*\theta((\eta_{\lvy\,*X})_{\eta_{\lvy} (w)}) \\
& = & \eta_{\lvy}(w)^{-1} (\pi_{Y\,\lvy*}\eta_{\lvy * X})_{\eta_{\lvy}(w)} \\
& = & \eta_{\lvy}(w)^{-1} \circ \eta_{Y*w}(\pi_{Y\,\lvy*} X) \\
& = & w^{-1} (\pi_{Y\,\lvy*}X) \\
& = & i^*\theta(X_w) \, .
\end{eqnarray*}
Conversely,
let $F\in\,\aut(\lvy)$ leave ${i^*\theta}$ invariant and cover $\eta_Y\in\,\aut Y$.
That is, $\pi_{Y\,\lvy}\circ F = \eta_Y \circ \pi_{Y\,\lvy}$.
Let $\tilde \eta_{\lvy}$ denote the natural lift of $\eta_Y$
and define $H := {\tilde \eta_{\lvy}}^{-1}\circ F$.  
Then $H$ leaves ${i^*\theta}$ invariant, and $\pi_{Y\, \lvy}\circ H = \pi_{Y\, \lvy}$. 
So for $X_w\in\,T_w(\lvy)$,
\begin{eqnarray*}
H^*i^*\theta(X_w) & = & i^*\theta(X_w) \\
H(w)^{-1}(\pi_{Y\,\lvy*}H_*X) &  = & w^{-1}(\pi_{Y\,\lvy*}X_w)     \\
H(w)^{-1}(\pi_{Y\,\lvy*}X)	& = & w^{-1}(\pi_{Y\,\lvy*}X)
\end{eqnarray*}
So $H(w) = w \;\,\forall w\in \lvy$ and thus $H=\mbox{Id}_{\lvy}$.~\blob

\bigskip

Denote the action of $\aut Y$ on $Y$ by
$\Psi:  \aut Y \,\times Y \rightarrow Y$.  
By Proposition~\ref{thm:liftlvy},
this action lifts to  an $(n+k)$--symplectic action 
$\tilde \Psi : \aut Y \times \lvy \rightarrow \lvy$.  
Now, a projectable  vector field
generates a  one-parameter group $\psi_t$ of  local automorphisms of $Y$,
which lifts to a one-parameter group $\tilde \psi_t$ of elements of 
 $\aut(U)$ for some open set $U \subset \lvy$.  
The vector field $\xi_U := (d/dt)\tilde \psi_t$
on $U$
 is the infinitesimal generator
of $\tilde \psi_t$ and thus is the natural lift
of $\xi_{\pi_{XY}(U)}$.

\begin{lemma}\label{lemma:Jlvy}
Let $\xi\in \projectable$ be the infinitesimal generator of the action of $\aut Y$
on $\lvy$.  Then
\[
\hat J (\xi) := \xi_{\lvy}\hook i^*\theta  
\]
is in $\tvlvy$, and its corresponding momentum mapping $J$
is $\operatorname{Ad}^*$ equivariant.
Furthermore, every element of $\tvlvy$ can be identified with 
$\hat J(\xi)$ for some $\xi \in \projectable$.
\end{lemma}

\proof Because $i^*\theta$ is invariant
under $\tilde\Psi$, it follows
 that $\lie_{\xi_{\lvy}}i^*\theta = 0$.  
Using~(\ref{eqn:liederiv}), we obtain
$d(\xi_{\lvy}\hook i^*\theta) = -\xi_{\lvy}\hook  i^*d\theta$ ,
so $\xi_{\lvy}\hook i^*\theta \in HF^1(\lvy)$.  
If $\tilde\psi_t$
is the local flow of $\xi_{\lvy}$ in a 
neighborhood of $w \in \lvy$,
 then for $g \in \ga$,
 $R_g\circ\tilde \psi_t = \tilde \psi_t \circ R_g$ 
 since $\tilde\psi_t$ is  a local 
 automorphism of $\lvy$.
Thus,  $R_{g*}\xi_{\lvy}(w) =\xi_{\lvy}(R_g(w))$.
Because $i^*\theta$ is $\ga$-tensorial, it now follows that  
$R^*_g(\xi_{\lvy}\hook i^*\theta)(w) =  g^{-1}\cdot (\xi_{\lvy}\hook i^*\theta)(w)$.
So $\xi_\lvy\hook i^*\theta \in \tvlvy$.
The proof that $J$ is $\operatorname{Ad}^*$ equivariant is exactly analogous to~\cite[Thm.\ 4.2.10]{AM}.
By Proposition~\ref{prop:bijective} any element of $\tvlvy$ can be identified with some
$\xi$ in $\projectable$ and thus $\hat J(\xi)$.~\blob

\bigskip

Let $\xi,\zeta \in \projectable$ and let $J$ be the momentum mapping 
induced by $\hat\Psi$.
Define an $\rnplusk$-valued Poisson bracket by 
\begin{equation}\label{eqn:tvlvypb}
\{\hat J (\xi), \hat J (\zeta) \}^\mu = -di^*\theta^\mu ( \xi_\lvy, \zeta_\lvy), \quad \mu = 1,2,\dots,n+k\; .
\end{equation}
By Proposition~\ref{thm:liftlvy}, identities~(\ref{eqn:liederiv}), and the
fact that $\pi_{*Y\,\lvy}[\xi_\lvy,\zeta_\lvy] = [\xi_Y,\zeta_Y]$,\begin{equation}\label{eqn:Jbracketlvy}
\{\hat J(\xi), \hat J(\zeta)\} = \hat J([\xi,\zeta]) \, .
\end{equation}
Observe that (\ref{eqn:Jbracketlvy}) has no additional term,
in contrast to~(\ref{eqn:Jbrackets}).

\section{Generalizing multisymplectic momentum mappings}
\setcounter{equation}{0}

Define a linear left action of $\ga $ on the vector space $\r^{n\times k}\times \r$ as follows.
\begin{equation}\label{eqn:ga-action}
(N,K,A)\cdot (B,\lambda) := \hbox{det}(N^{-1})\left(NBK^{-1}, \lambda - \tr(BK^{-1}A)\right)
\end{equation}
The associated vector bundle
$\lvy\times_{\ga } (\r^{n\times k}\times \r)$ 
is constructed  using the actions
in (\ref{eqn:galvy}) and (\ref{eqn:ga-action}).
For a point $\ptlvy\in \lvy$, define 
\begin{equation}\label{eqn:omega(e)}
\omega(e) := e^1\wedge e^2\wedge \cdots \wedge e^n
\qquad \mbox{and}\qquad
\omega(e)_i := e_i \hook \omega(e) \, ,
\end{equation}
where
$\{e^i\}$ is the dual basis of $\{e_i\}$.
From~(\ref{eqn:coframeaction}), 
if $(y,\{e_i',\epsilon_A'\})  = 
(y,\{e_i,\epsilon_A\})\cdot(N,K,A)$ then
$ \omega(e')  = \det (N^{-1})\omega(e) $
where $e = \{e_i\}$ and  $e' = \{e_i'\}$.

\begin{thm}\label{thm:z}
The map 
$$\begin{array}{rcl}
\hat\rho : \lvy\times_{\ga } (\r^{n\times k}\times \r) &
\rightarrow & \wedge^n Y \\
\left[ \ptlvy,(B,\lambda)\right] & \mapsto & (y, B^i_A\epsilon^A
\wedge\omega(e)_i + \lambda\omega(e))
\end{array}
$$
is a vector bundle monomorphism over $Y$.
The range of $\hat\rho$
is $Z$.
\end{thm}
 
\noindent 
The proof of the theorem appears in~\cite{La}.
Thus, the multisymplectic phase space $Z$
is a vector bundle over $Y$  associated to $\lvy$.
Also, the jet bundle $JY$ is associated to $\lvy$~\cite{La,MN}, but this construct is not necessary for us to proceed here.
From Theorem~\ref{thm:z}, the relationships between local
 coordinates $\{x^i,y^A,p^j_B,p\}$ on $Z$
and $\{x^i, y^A, \pi^j_k, \pi^B_C, \pi^D_l\}$ on
 $\lvy$ are
\[			
p^j_B = \det(\pi^a_b)\,B^i_A\pi^A_B(\pi^{-1})^j_i \; \quad\mbox{and}
\quad p = \det(\pi^a_b)(B^i_A\pi^A_k(\pi^{-1})^k_i + \lambda)\; .
\]			

Let $B\in\,\r^{n\times k}$ and $\lambda \in \r$. 
Define the map
\begin{equation}
  \phi_{(B,\lambda)} :  \lvy
  \rightarrow  Z : w \mapsto \hat\rho[w,(B,\lambda)]\, .
\end{equation}
The map  $\phi_{(B,\lambda)}$ preserves fibers over $Y$, and its range
is a subbundle of  $Z$
with standard fiber the $\ga$-orbit of $(B,\lambda)$.
The $\ga$-orbits for all nonzero $B \in \rnk$ are classified by the rank of $B$.

Using Proposition~\ref{prop:bijective} we may establish that 
for $\xi \in \projectable$, 
\begin{equation}\label{eqn:philvyz}
\phibl_*\xi_\lvy =  \xi_Z \, . 
\end{equation}
This is analogous to~\cite[Thm.\ 5.2]{No2} and is easily verified by local coordinate
calculations.

In order to derive 
the multisymplectic potential $\Theta$ on $Z$
from the $(n+k)$-symplectic potential 
$i^*\theta$ on $\lvy$,
we must make some preliminary remarks.
Let $\{R_\mu\}, \, \mu = 1,\dots,n+k$, be the standard basis of $\rnplusk$
and let $\{R^\mu\}$ be the corresponding dual basis.
Define the $\wedge^{r+s} \rnplusk$--valued $(p+q)$-form
$R_{\mu_1\cdots\mu_m} 
:= R_{\mu_1}\wedge\cdots\wedge R_{\mu_m} \in \wedge^m\rnplusk$
and 
$R^{\mu_1\cdots\mu_m} 
:= R^{\mu_1}\wedge\cdots\wedge
R^{\mu_m} \in \wedge^m{\rnplusk}^*$.
Let $\alpha$ be a $\wedge^r\rnplusk$-valued $p$-form and let $\beta$
be a $\wedge^s\rnplusk$-valued $q$-form on a manifold.  
Then 
$\alpha = \alpha^{\mu_1 \cdots \mu_r}\otimes R_{{\mu_1 \cdots \mu_r}}$ 
and
$\beta = \beta^{\mu_1 \cdots \mu_s}\otimes R_{{\mu_1 \cdots \mu_s}}$.
Define 
\[
\alpha\wedge\beta := (\alpha^{\mu_1 \cdots \mu_r}\wedge\beta^{\nu_1 \cdots \nu_s}
)\otimes R_{\mu_1 \cdots \mu_r\nu_1 \cdots \nu_s} \, .
\]
Let $\ptlvy \in \lvy$. 
From~(\ref{eqn:omega(e)}), 
\[				
\omega(e) := \frac{1}{n!}\epsilon_{i_1\cdots i_n}e^{i_1}\wedge\cdots\wedge e^{i_n}
\quad
\mbox{and}
\quad 
\omega(e)_j := \frac{1}{(n-1)!}\epsilon_{ji_1\cdots i_{n-1}}
e^{i_1}\wedge\cdots\wedge e^{i_{n-1}} \, ,
\]				
where $\epsilon_{i_1\dots i_n}$ is the sign of the permutation
$(1,\dots,n) \mapsto (i_1,\dots, i_n)$.

\begin{thm}\label{thm:nthetaV} 
For $n\geq 2$,
the $\wedge^n\rnplusk$-valued $n$-form $i^*\wedge^n\theta$ on $\lvy$
can be related to the canonical $n$-form $\Theta$ on $Z$ by 
\begin{equation}\label{eqn:relatedthetas}
\left< \wedge^n i^*\theta , \mbox{\boldmath $V$}(B,\lambda) \right> = \phi_{(B,\lambda)}^*\Theta
\end{equation}
where the map $\mbox{\boldmath $V$} : \rnk\times \r \rightarrow  \wedge^n{\r^{n+k}}^*$ has components 
\begin{eqnarray*}
V_{i_1\dots i_n}(B,\lambda) &=&\frac{1}{n!} \lambda\epsilon_{i_1\dots i_n}\, , \\
 V_{Ai_1\dots i_{n-1}}(B,\lambda)&=& \frac{1}{n!}B^j_{A}\epsilon_{ji_1\dots i_{n-1}}\\
 \mbox{and} \qquad V_{A_1\dots A_li_1\dots i_{n-l}}(B,\lambda)
      &=& 0 \;\; \forall \, l \geq 2.  
\end{eqnarray*}
\end{thm}

\noindent The theorem is proven in~\cite{La}.  Equation
(\ref{eqn:relatedthetas}) holds
for $n=1$ if $V_i(B,\lambda) = \lambda$ and $V_A(B,\lambda) = B_A$.

\begin{lemma}\label{lemma:Psis}
Let $\bar\Psi$ be the lift to $Z$ of the representation  
$\Psi$ of $\aut Y$ on $Y$.
Let $B\in \rnk$ and $\lambda \in \r$.
Then, using the identification $\lvy\times_{\ga}(\rnk\times \r) \simeq Z$
from Theorem~\ref{thm:z},
\[
\bar\Psi\left(\eta_Y,[\ptlvy,(B,\lambda)]\right) \simeq \left[\tilde\Psi(\eta_Y,\ptlvy),(B,\lambda)\right].
\]
As a consequence, 
$\phibl\circ \tilde\Psi_f(w) = \hat \Psi_f \circ \phibl(w)$.
\end{lemma}

\proof Using the action in (\ref{eqn:ga-action}) and Theorem~\ref{thm:z},
\begin{eqnarray*}
\left[\tilde\Psi(\eta_Y,\ptlvy),(B,\lambda)\right]
& \simeq & \left(\eta_Y(y),B^i_A(\eta_{Y*}\epsilon)^{*\,A}\wedge\omega(\eta_{Y*}e)_i
     + \lambda\omega(\eta_{Y*} e)\right) \\
& = & \left( \eta_Y(y),\eta_Y^{-1\,*}(B^i_A\epsilon^A\wedge \omega(e)_i 
	+ \lambda\omega(e))   \right) \\
& \simeq & \bar\Psi(\eta_Y, [\ptlvy, (B,\lambda)])\, .\;
\end{eqnarray*}
By choosing representatives of the equivalence classes, we see that
$\phibl\circ \tilde\Psi_f(w) = \hat \Psi_f \circ \phibl(w)$.~\blob

\bigskip

\begin{thm}\label{thm:phi} Let $w\in\lvy$ and $(B, \lambda) \in \rnk \times \r$.
$$
\phibl^*(\hat J_Z(\xi)) = \left<\hat J_{\lvy}(\xi)\wedge(\iwtheta {n-1}) , n \,\mbox{\boldmath $V$}(B, \lambda)\right>$$
where $\mbox{\boldmath $V$}(B, \lambda) \in \, \wedge^n\r^{(n+k)\,*}$ 
is defined in Theorem~\ref{thm:nthetaV}.
\end{thm}

\proof Let $w = \ptlvy\in \lvy$,
and let $\psi_t$ be a local one-parameter group of $X$.
Then $\tilde\psi_t$  is a flow on 
$\lvy$ and $\hat \psi_t$ is a flow on $Z$.  
It follows from Lemma~\ref{lemma:Psis} and equation~(\ref{eqn:philvyz}) that
$\phibl\circ\tilde\psi_t (w) = \hat \psi_t\circ \phibl (w)$ 
and thus
$\phibl_* \xi_{\lvy}(w) = \xi_Z (\phibl(w))$.
Using equation~(\ref{eqn:specialmommap}) and Theorem~\ref{thm:nthetaV},
\begin{eqnarray*}
\left< \hat J_{\lvy}(\xi)\wedge (\iwtheta {n-1}), n\, \mbox{\boldmath $V$}(B, \lambda)\right>
& = &  
    \left< (\xi_{\lvy}\hook \iwtheta {n}), \mbox{\boldmath $V$}(B, \lambda)\right>  \\
& = & \xi_{\lvy} \hook \phibl^* \Theta \\
& = & \phibl^*\left( \xi_Z\hook \Theta \right)   \\
& = & \phibl^*\hat J_Z(\xi) \, . ~\blob
\end{eqnarray*}

The last theorem  motivates us to represent
$\tvlvy$ in the space of 
$\wedge^{m+1}\rnplusk$-valued
$m$-forms by  
\begin{equation}\label{eqn:badrep}
\hat J(\xi)  \mapsto \hat J(\xi) \wedge(\iwtheta m) \, 
\end{equation}
for $0\leq m \leq n+k$. By induction, 
\begin{equation}\label{eqn:268}
\xi_{\lvy} \hook \iwtheta m = m\, \hat J(\xi) \wedge (\iwtheta {m-1})\,  .
\end{equation}
Because for each $0 \leq m \leq n + k -2$
the form $d(\iwtheta {m+1})$ is closed and nondegenerate,
 it is a candidate for a higher degree gen\-e\-ral\-i\-za\-tion of the
$(n+k)$-symplectic form $i^*d\theta$. 
Using (\ref{eqn:268}) and 
\[
d(\wedge^m\theta) = m\, d\theta\wedge(\wedge^{m-1}\theta) \, ,
\]
it follows that
\[
d(\hat J(\xi) \wedge(\iwtheta m))  = -\xi_\lvy \hook(i^*d\theta\wedge (\iwtheta {m}))\; .
\]
For $0 \leq m\leq n+k$ define a bracket on the image of   
representation~(\ref{eqn:badrep})
by
\[
\{\hat J(\xi) \wedge(\iwtheta m ) ,\hat J(\zeta) \wedge(\iwtheta m )\} : =
\xi_\lvy  \hook \zeta_\lvy \hook (i^*d\theta\wedge(\iwtheta m )) \, .
\]
If $m=0$ then this becomes the 
bracket on $\tvlvy$ given by~(\ref{eqn:tvlvypb}).
This sign convention for the bracket is consistent with~\cite{GIMMsy} 
and is the negative of that of~\cite{No1}. 
Computation on the forms yields, for $0 \leq m \leq n+k$,  
	\begin{equation}\label{eqn:fgtheta}
\{\hat f\wedge(\iwtheta m) ,\hat g\wedge(\iwtheta m) \}
= \{\hat f, \hat g\}\wedge(\iwtheta m) + m\,d(\hat f \wedge \hat g \wedge
(\iwtheta {m-1} ))
\, .
	\end{equation}

Equation~(\ref{eqn:fgtheta}) is
of particular interest when $m = n-1$, $\hat f = \hat J(\xi)$, and $\hat g = \hat J(\zeta)$.  
In this case we have reproduced  on $\lvy$ the exact analogue of the problem 
in~(\ref{eqn:Jbrackets}) that $T^1(Z)$ is not closed under its ``Poisson'' bracket. 
We could extend the algebra of momentum 
observables on $\lvy$ under representation~(\ref{eqn:badrep})  
to include closed $\rnplusk$-valued 
$(n-1)$-forms but its representation into $HV^1(\lvy)$ would have a 
kernel larger than that of the representation from
$\tvlvy$ into {\em the same} space of vector fields.
The space 
$\tvlvy$ of $\rnplusk$-valued Hamiltonian tensorial observables 
already possesses a well-defined Lie algebra under~(\ref{eqn:tvlvypb}),
Proposition~\ref{prop:bijective} shows that
 $\tvlvy$ and $T^1(Z)$ are in bijective correspondence,
and equation~(\ref{eqn:philvyz}) relates
the corresponding spaces of Hamiltonian  vector fields.  
Thus the $(n+k)$--symplectic geometry
of $\lvy$
not only generates the multisymplectic
geometry of $Z$ for classical fields but also possesses an algebraic structure that $Z$ lacks.

\section{Conserved quantities}
\setcounter{equation}{0}

Let $S_{\mu\nu}$ denote $R_\mu \otimes_s R_\nu$, where $\otimes_s$ is the symmetric tensor product.

\bigskip\noindent
{\bf Definitions}
 \begin{itemize}
	\item Let $ST^2(\lvy)$ denote the vector space of tensorial $\r^{n+k}\otimes_s\r^{n+k}$-valued functions on $\lvy$.
	
	\item Let $HF^2(\lvy)$ denote the vector space of $\r^{n+k}\otimes_s\r^{n+k}$-valued functions $\hat g = \hat g^{\mu\nu} S_{\mu\nu}$ on $\lvy$
that satisfy the symmetrized $(n+k)$-symplectic equation
	\begin{equation}\label{eqn:struc2}
d\hat g^{\mu\nu} = -2 X^{(\mu}\hook i^*d\theta^{\nu)}
	\end{equation}
where $ X^\mu R_\mu $ is an $\rnplusk$-valued vector field and parentheses denote symmetrization. 
	
	\item Let $HV^2(\lvy)$ denote the set of $\rnplusk$-valued vector fields  $ X_{\hat g}^\mu R_\mu $ satisfying equation (\ref{eqn:struc2}) for some $\hat g^{\mu\nu}\in HF^2(\lvy)$.  We shall call such vector fields \textit{Hamiltonian}.

	\item	Let $\stvlvy := ST^2(\lvy)\cap HF^2(\lvy)$.

\end{itemize}

\noindent Tensorial properties require that  $\hat g \in ST^2(\lvy)$ appear in local coordinates as
\begin{eqnarray}
\hat g(u) &=& g^{ij}\pi^k_i\pi^l_jS_{kl} + g^{ij}\pi^A_i\pi^l_jS_{lA} + g^{iA}\pi^l_i\pi^B_AS_{lB}
\\
&& + \; g^{ij}\pi^A_i\pi^B_jS_{AB} + g^{iC}\pi^A_i\pi^B_CS_{AB} + g^{CD}\pi^A_C\pi^B_DS_{AB} \, , \nonumber
\end{eqnarray}
where the component functions $g^{ij}, g^{iA}$, and $g^{AB}$ are functions of $x^k$ and $y^A$.  
If  $\hat g \in \stvlvy$ then the only additional restriction we have is that  
$g^{ij}= g^{ij}(x^k)$.
So $\stvlvy$ is in bijective correspondence with the space of projectable symmetric tensors of degree 2
on $Y$.  This is in exact analogy to Proposition~\ref{prop:bijective}.

As in the theory on $LM$~\cite{No1}, symmetrization in equation~(\ref{eqn:struc2}) 
means that the $\rnplusk$-valued Hamiltonian vector fields 
$X^\mu_{\hat g}R_\mu$ are not uniquely determined by $\hat g \in \stvlvy$.
Rather, they are determined locally up to the addition of vector fields
$Y^\mu R_\mu$ which satisfy
\begin{equation}\label{eqn:ymu}
Y^{(\mu} \hook i^*d\theta^{\nu)} = 0 \, .
\end{equation}
Thus each $Y^\mu$ must be a vertical vector field.  For a given
$\hat g \in \stvlvy$, two $\rnplusk$-valued Hamiltonian vector fields 
are in the same equivalence class
$\lb2 X_{\hat f}\rb2^\mu R_\mu$
 if their difference $Y^\mu$ satisfies~(\ref{eqn:ymu}).
Obtaining vector fields only up to equivalence
does not affect the basic algebraic structures of the $(n+k)$--symplectic geometry on
$\lvy$.  
For example, if $\hat f \in \tvlvy$ and $\hat g \in \stvlvy$,
define a Poisson bracket by 
\[
\{\hat f, \hat g\} := - X_{\hat f}(\hat g^{\mu\nu})S_{\mu\nu} 
\]
or use any representative of the equivalence class 
 $\lb2 X_{\hat g}\rb2^\mu$ to define a Poisson bracket
 by 
\[
\{\hat g, \hat f\} =  - 2 X_{\hat g}^{(\mu}(\hat f^{\nu)}) S_{\mu\nu}\, .
\]
The two are related by $\{\hat g, \hat f\} = - \{\hat f, \hat g\}$.
The algebra may be extended to higher degree observables and tensor-valued Hamiltonian vector fields, generating an algebra corresponding to the symmetric differential concomitants of Schouten~\cite{Sc} and Nijenhuis~\cite{Ni}.  See~\cite{No1,No2,No3} for details on how this is done on $LM$.  The proofs easily extend to $\lvy$.

\begin{lemma}\label{lemma:3.1}
	Let $\hat f = \hat f^\mu R_\mu \in \tvlvy$ and let $\xhat f$ be its corresponding Hamiltonian vector field. 
Let $\hat g\in \stvlvy$	and let $\xhat g$ be a representative of its equivalence class $\lb2\xhat g \rb2$ of corresponding $\rnplusk$-valued Hamiltonian vector fields. 
If $\pb g f = 0$, then for each $\mu=1,2,\dots,n+k$,  $\hat f^\mu$ is constant on the orbits of each $\xhat g^\mu \in \lb2\xhat g \rb2^\mu$.
\end{lemma}

\proof\cite{No2}  Let $F^\mu_t$ be the flow of $\xhat g^\mu \in \lb2\xhat g \rb2^\mu$.  
Then
\[
\frac{d}{dt}(\hat f^\mu \circ F^\mu_t)  
	=  (F^\mu_t)^* (\lie_{\xhat g^\mu}(\hat f^\mu))
	 =  (F^\mu_t)^* (\xhat g^\mu(\hat f^\mu)) 
	 =  - \frac{1}{2}(F^\mu_t)^* ({\pb g f}^{\mu\mu})\, . 
\]
So if $\pb g f = 0$ then $\frac{d}{dt}(\hat f^\mu \circ F^\mu_t) = 0$.  
This holds regardless of the choice of representative of  
$\lb2 \xhat g \rb2^\mu$.~\blob

\begin{thm}\label{thm:constmom}
Let $\Phi$ be an $(n+k)$-symplectic action of a subgroup $H$ of $\aut Y$ upon $\lvy$, and let $J$ be the corresponding momentum mapping.  Suppose that $\hat g \in \stvlvy$ is invariant under the action.
Then $J$ provides $n+k$ integrals of $\hat g$ in the sense that 
\[
J^\mu(F^\mu_t(u)) = J^\mu(u) \, ,
\]
where $F^\mu_t$ is the flow of any representative of $\lb2\xhat g \rb2^\mu$, 
$\mu = 1,2,\dots,n+k$.  
\end{thm}

\proof The proof follows~\cite[Appendix 1]{No2} and~\cite[Theorem 4.2.2]{AM}.
Let $\lieh$ be the Lie algebra of $H$.
From our remarks in Section~\ref{sec:multisymplectic}, 
if $H \subset \aut Y$ then $\lieh \subset \projectable$.  
For each $\xi$ in $\lieh$, 
invariance of $\hat g$ under $\Phi$ implies
\begin{equation}\label{eqn:ginvariant}
\hat g (\Phi_{\exp(t\xi)}(u)) = \hat g (u) \, .
\end{equation}
 Differentiating~(\ref{eqn:ginvariant}) at $t=0$ gives us 
$d\hat g (\xi_{\lvy})= 0 \, ,
$
which implies that 
\begin{equation}\label{eqn:LieJ}
\lie_{\xi_\lvy}(\hat g) = 0 \, .
\end{equation}
By Lemma~\ref{lemma:Jlvy} 
if $\xi \in \projectable$, then $\hat J(\xi) \in \tvlvy$.
This and (\ref{eqn:LieJ}) imply that 
\[
\{\hat J(\xi),\hat g\} = 0
\, .
\]
 Hence, by Lemma~\ref{lemma:3.1}, 
\[
(\hat J (\xi))^\mu(F^\mu_t(u)) = (\hat J(\xi))^\mu(u) \, .
\]
Thus, by the definition of a momentum mapping on $\lvy$,
\[
J^\mu(F^\mu_t(u)) = J^\mu (u) \, .~\blob
\]

\section{Examples}

\setcounter{equation}{0}

\subsection{Linear momentum in field theories}

Following examples in~\cite{AM} and~\cite{No2},
we express linear momentum 
in the language of momentum mappings on $\lvy$ and on $Z$.
Let $X = \r^n$, $Y = \rnplusk$, and $\lieG = (\rnplusk, +)$.
(So $\lieg$ is the abelian Lie algebra $\rnplusk$.)
Define an action of $\lieG$ on $Y$ by 
\begin{equation}
\Psi: \lieG\times Y \rightarrow Y : (g,y) \mapsto \Psi_g (y) := g + y\, .
\end{equation}
Define 
\begin{equation}
\underline\Psi : \lieG\times X \rightarrow X : (g,x)\mapsto
\underline\Psi_{g}(x) := \operatorname{Pr}_1(g) + x \, ,
\end{equation}
where $\operatorname{Pr}_1:\rnplusk \rightarrow \r^n$ is projection onto the first factor.
Then $\Psi_g$ and  $\underline \Psi_g$ are bijective and
$\underline\Psi_{g}\circ \pi =  \pi\circ \Psi_g$.
Thus $g\mapsto \Psi_g$ is a representation of $\lieG$ in $\aut(Y)$.  
Now we may compute the generators of the action.
First, define a lift $\tilde\Psi$ of the representation $\Psi$ to  $\lvy$ by
\begin{eqnarray}
\tilde\Psi(g,\ptlvy) & = & \left(  \Psi_g (y), \{{\Psi}_{g*}e_i,\Psi_{g*}\epsilon_A\}\right) \nonumber \\
& = & (y + g, \{e_i,\epsilon_A\}).
\end{eqnarray}
Let $\xi\in \lieg$.  
The infinitesimal generator on the action on $\lvy$ is 
	\begin{equation}\label{eqn:infgenlvy}
\xi_{\lvy}\ptlvy 
 =  \left. \frac{d}{dt} (y + \exp tX, \{e_i,\epsilon _A\} )
 \right|_{t=0}\, . 
	\end{equation}
If $\xi_{\lvy} = \xi^i\hat r_i + \xi^A\hat s_A$ then 
we may write equation~(\ref{eqn:infgenlvy})
locally on $\lvy$ as 
\begin{equation}\label{eqn:xilvy}
\xi_{\lvy} = \xi^i \,\basisx i +  \xi^A \,\basisy A \, .
\end{equation}
From equations~(\ref{eqn:idtheta}) and~(\ref{eqn:xilvy}),
the corresponding momentum mapping is 
\begin{equation}\label{eqn:linPilvy}
\hat J_{\lvy}(\xi)  =  \xi_{\lvy}\hook{i^*\theta}
 =  \xi^i\pi^j_i\hat r_j + \xi^A\pi^B_A\hat s_B + \xi^i\pi^B_i\hat s_B 
\end{equation}
or
\[
J_{\lvy} = \pi^j_i\hat r^i\otimes \hat r_j + \pi^A_B\hat s^B\otimes\hat s_A
       + \pi^B_i\hat r^i\otimes \hat s_B.
\]
The $\rnplusk$-valued function $\hat J_{\lvy}(\xi)$ is interpreted as a {\em momentum frame}
~\cite{No2}.

Now consider linear momentum on $Z$, 
using the same $\lieG$ and $Y$ as before.
The lift of $\Psi$ to $Z$ is given by 
\[
\hat{\Psi_g}(z) = \Psi^{-1\,*}_g \,z \, .
\]
As with the case on $\lvy$,  in local coordinates,
\[
\xi_Z(z) =  \xi^i\basisx i + \xi^A\basisy A.
\]
On Z we may write
\[
\hat J_Z(\xi) = \xi_Z\hook\Theta\, ,
\]
or locally,
\begin{equation} \label{eqn:linPZ} 
\hat J_Z(\xi) 
 = 
\xi^Ap^i_A(d^{n-1}x)_i + \xi^ip^j_A dy^A \wedge(d^{n-2}x)_{ij} + \xi^i p  (d^{n-1}x)_i \, ,
\end{equation}
which has the interpretation of the linear momentum of a field and
is consistent with~\cite[p.\ 45]{GIMMsy}.
To verify Theorem~\ref{thm:phi}, namely that $\left<\hat J_{\lvy}(\xi)\wedge(\iwtheta {n-1}) , \mbox{\boldmath $V$}\right>$ $= \phibl^*(\hat J_Z(\xi))$ ,
we may perform a local coordinate calculation 
comparing~(\ref{eqn:linPZ}) with~(\ref{eqn:linPilvy}).

\subsection{Angular momentum in field theories}

Motivated by examples on the cotangent bundle~\cite{AM} 
and on the linear frame bundle~\cite{No2},
we express angular momentum 
in the language of momentum mappings on $\lvy$ and on $Z$.
Let $X = \r^n$ and $Y = \rnplusk$
and let $\{x^i,y^A\}$ be adapted coordinates on $Y$. 
Let $\eta$ be a constant nondegenerate metric on $\r^n$ 
and let $\iota$ be a constant nondegenerate metric on
$\r^k$. 
(For example, to model the space of $4$-vector electromagnetic potentials over Minkowski spacetime, both
 $\eta$ and $\iota$ could be the Lorentz metric on $\r^4$.)
Construct the \textit{Kaluza-Klein metric} $G$ on $Y$ using 
an Ehresmann connection $\gamma: TY\rightarrow V(TY)$.
The Kaluza-Klein metric is the fiberwise bilinear map
G on $TY$ given by 
\[
G(v,w) = \pi_{XY}^*\eta(v,w) + \iota(\gamma(v), \gamma(w)) \, .
\]
If we use the trivial Ehresmann connection
$\gamma: \rnplusk\rightarrow \r^n: v^i\hat r_i + v^A\hat s_A \mapsto v^A$, then
$G$ is itself constant and nondegenerate, $G|\gamma(TX) =\eta$,
and $G|V(TY)=\iota$.  In adapted coordinates,
\[
G(v,w) = \eta_{ij}v^iw^j + \iota_{AB}v^Aw^B \, .
\]

Define a
left action $\Phi :(\onk)\times Y \rightarrow Y$ by left matrix multiplication
on column vectors in $\rnplusk$, and $\Phi_g(y) := \Phi(g,y)$.
Define a left action  $$\underline\Phi : (\onk) \times X\rightarrow X$$
by $\underline\Phi (g,x) = \pi\circ \Phi(g\,\hat x) =:\underline\Phi_g(x)$ where
 $\r^n \hookrightarrow \rnplusk: x \mapsto\hat x$ is inclusion. 
Observe that $\Phi_g$ and $\underline\Phi_g$ are bijections and
that $\underline\Phi_g \circ \pi = \pi \circ \Phi_g$.  Thus
$g \mapsto \Phi_g$ is a representation of $\onk$ in $\aut(Y)$.

\rem The larger group $O_G(n+k)$ does not possess
a nontrivial left action on $X=\r^n$.
Under left multiplication of $GL(n+k)$ on $\rnplusk$,
  $g\in GL(n+k)$  leaves $\r^n$ invariant
if and only if $g \in G_A$. 
Thus, we use $\onk = G_A \cap  O_G(n+k)$.

\bigskip

To compute the infinitesimal generators, 
let $\xi = \xi^\mu_\nu E^\nu_\mu \in \onkalg$,
where $\{E^\nu_\mu\}$ is the standard basis for 
the Lie algebra $\mathfrak{gl}(n+k)$ and the subset 
$\{E^i_j, E^A_B\}$ is a basis for $\onkalg$.
 Then the infinitesimal generator 
of the action of $\xi$ on $Y$  is 
\[
\xi_Y(y) = \xi^i_j x^j \basisx i + \xi^A_B y^B \basisy A \; .
\]
Note that $\pi_{XY*}(\xi_Y(y)) = \xi^i_j x^j \basisx i  = \xi_X(\pi_{XY*}(y))$.
Observe that $\xi_Y$ is a Killing vector field of $G$.
Indeed it is easy to verify that 
\begin{equation}\label{eqn:Killing}
G_{\mu\nu} (\xi_Y)^{\nu},_{\kappa} + G_{\kappa\nu}(\xi_Y)^{\nu},_{\mu}=0\, .
\end{equation}
If $\kappa = C$ and $\mu = D$, then the left side of (\ref{eqn:Killing}) simplifies to
$\iota_{DA}\xi^A_C +\xi^A_D \iota_{AC}$, 
which vanishes because $(\xi^A_B) \in {\cal O}_\iota(k)$.
If $\kappa = k$ and $\mu = l$, then the left side of (\ref{eqn:Killing})
becomes $\eta_{li}\xi^i_k +  \xi^i_l \eta_{ik}$,
which vanishes because $(\xi^i_j) \in {\cal O}\eta(n)$.
If $\kappa = l$ and $\mu = C$ or vice-versa, the 
left side of (\ref{eqn:Killing}) is identically
zero since $G_{iA} = 0$.

Let $u \in \lvy$ and $\xi \in \onkalg$.  The momentum mapping on $\lvy$ associated
with the action defined above is
\[
\hat J_{\lvy} (\xi)(u) = \xi^j_kx^k\pi^i_j \hat r_i + 
(\xi^j_kx^k\pi^A_j + \xi^B_Cy^C\pi^A_B)\hat s_A \, ,
\]
and if  $(C^\mu_\nu)$ is the basis of $\mathfrak{gl}(n+k)^*$ dual
to $(E^\nu_\mu)$ then  
\begin{eqnarray}
J(u) &=&  C^j_kx^k\pi^i_j \hat r_i + 
(C^j_kx^k\pi^A_j + C^B_Dy^D\pi^A_B)\hat s_A \nonumber\\
& = & 
\eta_{kl}C^{lj}x^k\pi^i_j \hat r_i + 
(\eta_{kl}C^{jl}x^k\pi^A_j + \iota_{DE}C^{DB}y^E\pi^A_B)\hat s_A  \\
& = & \frac{1}{2}\left(
(x_i\pi^k_j -x_j\pi^k_i)C^{ji}\hat r_k + \left(
(x_i\pi^A_j - x_j\pi^A_i)C^{ji} + (y_D\pi^A_B - y_B\pi^A_D)C^{BD}
\right)\hat s_A
\right)\; .
\nonumber
\end{eqnarray}
The function $\hat G$ corresponding to the metric $G$  is in $\stvlvy$,
and via equation~(\ref{eqn:struc2}),  produces 
a system of $n+k$ Hamiltonian
vector fields,
\begin{equation}
\left\{
\begin{array}{rcl}
X^i_{\hat G} & = & \eta^{jk}\pi^i_j\basisx k \\
X^A_{\hat G} & = & \eta^{jk}\pi^A_j\basisx k + \iota^{BC}\pi^A_B\basisy C \; .  
\end{array}
\right.
\end{equation}
Equation~(\ref{eqn:Killing}) implies that the $X^\mu_{\hat G}$ are tangent to the $\on \times \ok$ subbundle of $\lvy$.
By Theorem~\ref{thm:constmom},
each $J^\mu$ is constant along $X^\mu_{\hat G}$ for the same $\mu$.
We may interpret $J^i$ to be the $i^{\mbox{\tiny th}}$ component of the (extrinsic)
angular momentum in parameter space
(e.g., Minkowski spacetime) and $J^A$ to be the 
$i^{\mbox{\tiny th}}$ component of the total angular momentum
(both extrinsic and field) in the field configuration space.
As a result, along an integral curve (a solution of Hamilton's equations for the
metric Hamiltonian),
the $X^i_{\hat G}$ gives the equations
$\dot x^k = \eta^{kj}\pi^i_j$ and  $\dot y^A = \dot \pi^\mu_\nu = 0$.
Integrating with an initial condition $u_0 = (y_0,\{e_i,\epsilon_A\}_0)$, we obtain in local coordinates
the flow
$F^i_t =(x^k + t\eta^{kj}\pi^i_j, y^A, \pi^m_l,\pi^B_n,
\pi^C_D)$, which is  the same as the flow of $X_{\hat g}^i$ on $X$.
In fact,
\begin{eqnarray}
J^i \circ F^i_t &=& \pi^i_j(x^k + t\, \eta^{kl}\pi^i_l)C^j_k \qquad \mbox{(no sum on $i$)}\nonumber \\
& = &  \pi^i_jx^kC^j_k + t\pi^i_j\pi^i_lC^{jl}\nonumber \\
& = & \pi^i_jx^kC^j_k \nonumber\\
& = & g_{jl}\dot x^l x^kC^j_k \nonumber\\
& = & \dot x^{[l} x^{k]} C_{lk} \nonumber\\
& = & J^i \, .\label{eqn:Ji}
\end{eqnarray}
From the $X^A_{\hat G}$ we obtain
$\dot x^i = \eta^{ij}\pi^A_j, \dot y^C = \iota^{BC}\pi^A_B$, and $\dot \pi^\mu_\nu = 0$.
The  flow for  $X^A_{\hat G}$ is 
\[
F^A_t = (x^k + t\eta^{kj}\pi^A_j, y^C + t \iota^{CB}\pi^A_B, \pi^l_m,\pi^D_n,
\pi^E_F) \, ,
\]
and
\begin{eqnarray}
J^A \circ F^A_t &=& \pi^A_i(x^k + t\, \eta^{kl}\pi^A_l)C^i_k + 
\pi^A_D(y^B + t\,\iota^{BE}\pi^A_E)C^D_B \qquad \mbox{(no sum on $A$)} \nonumber \\
& = & 
\pi^A_jx^kC^j_k + \pi^A_By^DC^B_D + t\pi^A_i\pi^A_lC^{il} +
 t\pi^A_D\pi^A_BC^{DB} \nonumber \\
& = & 
\pi^A_jx^kC^j_k + \pi^A_By^DC^B_D \nonumber\\
& = & \dot x^{[l} x^{k]} C_{lk} + \dot y^{[D} y^{B]} C_{DB} \nonumber\\
& = & J^A \, . \label{eqn:JA}
\end{eqnarray}

\noindent Thus along a trajectory (integral curve)
extrinsic angular momentum is conserved (equation~(\ref{eqn:Ji})) and
total angular momentum is conserved (equation~(\ref{eqn:JA})).

If $n = 1$ then we may model time-evolved particle mechanics (with $k$ degrees of 
freedom).  The group action is, up to time rescaling, merely that of the orthogonal group in the $k$ spatial dimensions.
Equation~(\ref{eqn:Ji}) reduces to $J^i\circ F^i_t \equiv 0$, and 
equation~(\ref{eqn:JA}) reduces to $J^C\circ F^A_t = \dot y^{[A} y^{B]} C_{AB}\,$.
Thus time parametrization has no effect on the conservation of
angular momentum.
If $k = 1$ then we may model scalar fields over parameter space.  
The group action is, up to rescaling in the field, merely that of the orthogonal group in the $n$ spatial dimensions. 
Then equation~(\ref{eqn:Ji}) is left intact and 
equation~(\ref{eqn:JA}) reduces to   equation~(\ref{eqn:Ji}).  The scalar field angular momentum
comes purely from the parameter space.

\subsection{Affine symmetry in time-evolution mechanics}

Let $X = \r$ and 
let $Y = \r \times \r^k$,
with coordinates $\{x^0=t,y^A\}$.
Give $X$ the Euclidean metric, and let 
$\iota$ be a constant nondegenerate metric 
on the fiber $\r^k$.
Assume the trivial connection  of $Y$ over $X$ to construct the 
Kaluza-Klein metric $G$ on $Y$,
\[
G(v,w) =  v^0w^0 + \iota_{AB}v^Aw^B \, .
\]

Construct a subgroup  of $GL(k+1)$ by
\[
{\lieG}(k):= \left\{ \left. \left(
\begin{array}{cc}
1  & 0 \\ v & K
\end{array}\, \right)  
\right| \, K\in \ok, \, v\in \r^k
 \right\} \, .
\]
The group $\lieG (k)$ is the semidirect product of $\r^k$ with $\ok$.
The left representation $\Phi: \lieG (k) \times Y \rightarrow Y$ defined 
by left matrix multiplication covers the trivial representation of 
$\lieG (k)$ on  $X=\r$,
so that for all $g \in \lieG (k)$, $\Phi_g \in \aut (Y)$.
Lift $\Phi_g$ to an action of $\lieG(k)$ on $\lvy$.
The corresponding Lie algebra $\lieg(k)$
is the semidirect product of $\r^k$ with $\okalg$,
and may be represented  as a subalgebra of $\mathfrak{gl}(k+1)$
by
\[
\lieg(k) = \left\{ \left. \left(
\begin{array}{cc}
0 & 0 \\
 v & k
\end{array}\, \right)
\right| \, k\in \okalg, \, v\in \r^k
  \right\}
\]
If we express the basis of $\lieg(k)$ by $\{E^A_B , s_C\}$
where $\{E^A_B \}$ is the standard basis for $\okalg$
and $\{s_C\}$ is the standard basis for $\r^k$,
For $\xi \in \lieg(k)$, we may write $\xi = k^A_BE^B_A + v^A s_A$. 
In local coordinates the corresponding infinitesimal generator on $Y$ is
$\xi_Y(y) = (k^A_By^B + x^0v^A)\basisy A $.
The corresponding momentum mapping on $\lvy$ is
\[
\hat J(\xi)(u) = (k^A_Cy^C\pi^B_A + x^0v^A\pi^B_A)\hat s_B
\]
or if $\{C^A_B, s^D\}$ is the basis of $\lieg(k)^*$ dual to $\{E^A_B , s_C\}$,
\[
J(u)  =   (y^D\pi^B_AC^A_D + x^0\pi^B_As^A)\hat s_B
 =   (y_{[D} \pi^B_{A]} C^{AD} + x^0\pi^B_As^A)\hat s_B \; .
\]

The function $\hat G$ corresponding to the metric $G$  is in $\stvlvy$
and, via equation~(\ref{eqn:struc2}),  produces 
an equivalence class  of $\r^{1+k}$-valued Hamiltonian vector fields 
$\lb2 X_{\hat G}\rb2$.  
We choose a representative of this equivalence class, expressed in local coordinates as
\[
X^0_{\hat G} = \pi^0_0 \frac{\partial}{\partial x^0} 
\quad \mbox{and} \quad
X^A_{\hat G}  =   \pi^A_0 \frac{\partial}{\partial x^0}  + 
\iota^{BC}\pi^A_B\basisy C
 \, .
\]
Along an integral curve of $X^0_{\hat G}$,
we get the equations $\dot x^0 =  \pi^0_0$ and 
$\dot y^A = \dot\pi^B_C = \dot\pi^D_0 = \dot\pi^0_0 =0$.
Integrating with suitable initial conditions we obtain
$F^0_\lambda = (x^0 + \pi^0_{0}\lambda,
y^A,\pi^B_{C},\pi^D_{0},\pi^0_{0})$.
So,
\[
J^0\circ F^0_\lambda = J^0 = 0 \, .
\]
Along an integral curve of $X^A_{\hat G}$,
we obtain the equations 
\[
\dot x^0  =   \pi^A_0 \, , \qquad  
\dot y^C   = \iota^{BC}\pi^A_B \, , 
	\quad \mbox{and} \quad
\dot\pi^B_C  =   \dot\pi^D_0 = \dot\pi^0_0 =0 \, . 
\]
Integrating with initial conditions we get
$F^A_\lambda = (x^0 + \pi^A_{0}\lambda,
y^C + \lambda\iota^{BC}\pi^A_{B} ,
 \pi^B_{D},\pi^E_{0},\pi^0_{0})$.
So,
\begin{eqnarray}
J^B \circ F^B_\lambda  
	& = &(x^0 + \pi^B_0 \lambda)\pi^B_As^A +
 (y^C + \lambda\iota^{CE}\pi^B_C )\pi^B_D C^D_E  \qquad\mbox{no sum on $B$} \nonumber \\
	& = & x^0\dot y^C \iota_{AC}s^A + y^E\dot y^D \iota_{AD}C^A_E + \lambda \dot x^0
\iota_{AC} \dot y^C s^A \nonumber\\
	& = &  y_{[E} \dot y_{D]} C^{DE}  
+ x^0\dot y^C \iota_{AC}s^A + \lambda \dot x^0
\iota_{AC} \dot y^C s^A  \nonumber\\
	& = & J^B +  \lambda \dot x^0
\iota_{AC} \dot y^C s^A
\; .  \label{eqn:JB}
\end{eqnarray}

Apparently (\ref{eqn:JB}) violates conservation of angular momentum along the flow, because
$F^B_\lambda$ is \textit{not} an integral of the motion.  
Closer observation reveals
that the
additional term describes the contribution to the angular momentum from
shift of the point in $Y$ under the translational part  of $\lieG (k)$.
Indeed if we restrict the action to 
$\ok \subset \lieG (k)$, then the extra term disappears as in 
equation~(\ref{eqn:JA}), and we produce 
conserved angular momentum along flows in the spatial directions.
Thus the $(n+k)$--symplectic geometry adjusts the angular momentum
to accommodate a change in reference frame.

If $k=3$ and $\iota$ is the Euclidean metric on $\r^3$, then $\lieG(3) = E(3)$, the Euclidean group.  Equation
(\ref{eqn:JB}) may be interpreted to be the ``parallel axis theorem''
of classical mechanics.  If $k=4$ and $\iota$ is the Lorentz metric, then $\lieG(4) = E(4)$, the Poincar\'e group, and 
(\ref{eqn:JB}) indicates how to transform $4$-angular momentum when boosting from one
relativistic inertial reference frame to another.
	
\subsection{Time reparametrization symmetry in time-evolution mechanics}

Let $X = \r$ and let $Y = \r \times Q$, where $Q$ is a $k$-dimensional manifold. 
Again, coordinates on $y$ are $\{x^0=t,y^A\}$.
Let $X$ have the Euclidean metric and let $Q$ have a Riemannian metric $g$.
Assume a generic connection $\gamma: TY \rightarrow V(TY)$ expressed locally by
$\gamma(v) = (v^A-v^0\gamma^A)\basisy A$.
The resulting
Kaluza-Klein metric $G$ on $Y$ has local coordinate expression,
\[
G(v,w) =  (1+g_{AB}\gamma^A \gamma^B)v^0w^0 
+ g_{AB}\gamma^A(v^Bw^0 + v^0w^B) 
+ g_{AB} v^Aw^B \, .
\]

Let $\lieG = \diff \r$ act on $Y$ by time reparametrizations.
The corresponding Lie algebra $\lieg$ may be identified with
$C^\infty(\r)$, and we may identify $\lieg^*$, the space of densities on $\r$, with $\lieg$. 
This action can be represented in $\aut Y$, as it leaves the fiber of $Y$ constant.  
Clearly this action may be lifted to $\lvy$.
In local coordinates, $f\in \diff \r$ has infinitesimal generator
$f_Y = f(x^0) \frac{\partial}{\partial x^0}$,
which lifts to a projectable vector field on $\lvy$,
$f_\lvy = f(x^0) \frac{\partial}{\partial x^0}$.
Using Lemma \ref{lemma:Jlvy} the corresponding momentum observable
is 
\[ 				
\hat J_\lvy (f)  
= f(x^0) \frac{\partial}{\partial x^0} \hook ((\pi^A_Bdy^B + \pi^A_0dx^0)\hat s_A
 +\pi_0^0dx^0\hat r_0)
= f(x^0)(\pi^A_0\hat s_A + \pi^0_0 \hat r_0)
\]				
or 
\[
J_\lvy : f \mapsto f \cdot (\pi^A_0\hat s_A + \pi^0_0 \hat r_0) \, .
\]
On $\lvy$ we may note the time-dependence of the momentum observables.  In fact the second term describes momentum along the time parameter and the first term provides an adjustment in momentum in the fiber of $Y$ to account for the time reparametrization.

The lift of the action to $Z$ is 
$ f_Z = f_Y = f(x^0) \frac{\partial}{\partial x^0}$,
and its momentum observable on $Z$ is 
\[
\hat J_Z (f)(x^0) = f_Z \hook \Theta = p f(x^0) \, . 
\]
Note the time-dependence of the momentum observables
on $Z \simeq T^*\r \times T^*Q$.

The tensorial function $\hat G$ on $\lvy$ 
corresponding to the Kaluza-Klein metric $G$ 
is in $\stvlvy$
and, by~(\ref{eqn:struc2}), produces equivalence classes
of Hamiltonian vector fields $\lb2 X_{\hat G} \rb2^\mu$.
If in addition to~(\ref{eqn:struc2}) the Hamiltonian vector fields must satisfy a ``no-torsion'' condition,
\[
X_{\hat g}^\mu \hook X_{\hat g}^\nu \hook i^*d\theta = 0 \, ,
\]
then the Hamiltonian vector fields are unique and may be 
expressed locally as 
\begin{equation}\label{eqn:xgmu}
X_{\hat g}^\mu =  G^{\nu\lambda}\pi^\mu_\nu \frac{\partial}{\partial Y^\lambda}
	+ \Gamma^\nu_{\sigma\rho} G^{\kappa\rho} \pi^\mu_\kappa \pi^\lambda_\nu
		\frac{\partial}{\partial \pi^\lambda_\sigma}
\end{equation}
where $\{Y^\mu\} = \{x^0,y^A\}$ and 
$\Gamma^\mu_{\nu\lambda}$ are the Christoffel symbols of the 
second kind for the Levi-Civita connection defined by $G$.
Note that several terms such as $\pi^i_A, \Gamma^0_{0i}$, and $\Gamma^k_{00}$
will vanish in~(\ref{eqn:xgmu}).
From this we may write the differential equations for the integral
curve of $X_{\hat g}^0$ through a point $u\in\lvy$.
We get (among others) a geodesic equation:
\[
\ddot{y}^A + \Gamma^A_{BC}\dot y^B\dot y^C 
	  + 2\Gamma^A_{B0}\dot y^B\dot x^0 
	+ \Gamma^A_{00}\dot x^0\dot x^0  
 = 0 \, ,
\]
and an equation of parallel transport of vertical frames,
\[
\ddot{\pi}^A_B - \Gamma^C_{BD} \dot y^D \pi^A_C  - \Gamma^0_{BD} \dot y^D \pi^A_0
			- \Gamma^C_{B0} \dot x^0 \pi^A_C - \Gamma^0_{B0} \dot x^0 \pi^A_0
	  = 0 \, .
\]
This improves upon a ``parallel transport'' result of Norris~\cite{No2,No3}, 
because for a nontrivial connection $\gamma$ we obtain parallel transport of frames along the time direction without needing to assume that one of the legs of the frame is tangent to the geodesic.

\section{Conclusions}

\setcounter{equation}{0}

This investigation 
shows that the
$(n+k)$-symplectic geometry of the bundle
of vertically adapted
linear frames $\lvy$ of a field configuration bundle $Y$
extends the $n$-symplectic geometry of Norris~\cite{No1,No2,No3}
to provide momentum mappings for field theories.
The``allowable'' classical field momentum observables are a special case of these momentum mappings,
generated by lifting a bundle 
automorphism of $Y$ to a momentum mapping on $\lvy$. 
These momentum observables improve upon the  momentum observables found in the multisymplectic geometry of Gotay, et al.~\cite{Go,GIMMsy}.
Furthermore, we may interpret a Kaluza-Klein metric tensor on $Y$ 
as the energy observable of the system and identify conserved quantities as those whose
corresponding observables commute with the energy observable.  
Examples demonstrate conservation of the 
field, parameter space, and total momentum.
Furthermore, the frame bundle perspective can allow for adjustments in the conservation laws to accommodate a change in inertial reference frame.

This work sets the stage for an $(n+k)$-symplectic reduction by symmetry.  Reduction by symmetry in this context would address field theories with symmetry but would employ a finite-dimensional approach.
Progress on the Lagrangian side of this problem is seen using the multisymplectic
approach~\cite{CRS,MPS}
and in the $n$-symplectic approach~\cite{MN,No4}.   It is 
hoped that a full program of field-theoretic reduction by symmetry, analogous to cotangent bundle, Lie-Poisson,  or Euler-Poincar\'e reduction, will emerge in the recent future.

\section*{\bf Acknowledgements} 
This research was partially  supported by ROA Supplements to NSF grants DMS 9802106 and DMS 9633161 at the California Institute of Technology and by the John M. Bennett, Sr., Fellowship at Trinity University.  
The author thanks J. E. Marsden for his assistance with obtaining financial support and for his comments.
The author also thanks R. O. Fulp and L. K. Norris for their insights.


\end{document}